\documentstyle[12pt]{article}

\begin{document}

\baselineskip 22pt



\begin{center}

{\Large \bf Dependence of $|V_{ub}/V_{cb}|$\\
on Fermi momentum $p_{_F}$ in ACCMM model
\\}

\vspace{1.0cm}
Dae Sung Hwang$^1$, C.S. Kim$^2$ and Wuk Namgung$^3$\\

$1$: Department of Physics, Sejong University, Seoul 133--747, Korea\\
$2$: Department of Physics, Yonsei University, Seoul 120--749, Korea\\
$3$: Department of Physics, Dongguk University, Seoul 100--715, Korea\\

\vspace{2.0cm}

{\bf Abstract} \\

\end{center}

The Gaussian width of Fermi momentum, $p_{_F}$, is the most important
parameter of the ACCMM model, and
its value is essential in the determination of $|V_{ub}/V_{cb}|$
because the experimental analysis is allowed only at the end-point region of
inclusive semileptonic $B$-decay spectrum.
We extract the value of $|V_{ub}/V_{cb}|$ as a function of $p_{_F}$.
We also calculate the parameter $p_{_F}$ in the
relativistic quark model using the variational method,
and obtain $p_{_F} = 0.54$ GeV which is much larger than the commonly used
value, $\sim 0.3$ GeV, in experimental analyses.
When we use $p_{_F} = 0.5$ GeV instead of 0.3 GeV,
the value of $|V_{ub}/V_{cb}|$ from ACCMM model is increased by a factor 1.81,
and can give a good agreement with Isgur {\it et al.} model.

\vfill

\noindent
$1$: e-mail: dshwang@phy.sejong.ac.kr\\
$2$: e-mail: kim@cskim.yonsei.ac.kr
\thispagestyle{empty}
\pagebreak

\baselineskip 22pt

\noindent
{\bf \large 1. Introduction}\\

In the minimal standard model CP violation is possible through the CKM mixing
matrix of three families, and it is important to know whether the element
$V_{ub}$ is non-zero or not accurately. Its knowledge is also necessary to
check whether the unitarity
triangle  is closed or not \cite{quinn}.
However, its experimental value is very poorly known presently and its better
experimental information is urgently required. At present, the only
experimental method
to measure $V_{ub}$ is through the end-point lepton
energy spectrum of the inclusive $B$-meson
semileptonic decays, {\it e.g.} CLEO \cite{cleo} and ARGUS \cite{argus},
and their data indicate that $V_{ub}$ is non-zero.
Recently it has also been  suggested  that the measurements of
hadronic invariant mass spectrum \cite{kim} as well as
hadronic energy spectrum \cite{bouzas}
in the inclusive $B \rightarrow X_{c(u)} l \nu$ decays can be
useful in extracting $|V_{ub}|$ with better theoretical understandings.
In future asymmetric $B$ factories with vertex detector, they will offer
alternative ways to select $b \rightarrow u$ transitions that are much more
efficient than selecting the upper end of the lepton energy spectrum.

The simplest model for the semileptonic $B$-decay is the spectator model which
considers the decaying $b$-quark in the $B$-meson as a free particle.
The spectator model
is usually used with the inclusion of perturbative QCD
radiative corrections \cite{kuhn}.
Then the decay
width of the process $B\rightarrow X_ql\nu$ is given by
\begin{equation}
{\Gamma}_B (B\rightarrow X_ql\nu )\simeq {\Gamma}_b (b\rightarrow ql\nu )=
\vert V_{bq}{\vert}^2({{G_F^2m_b^5}\over {192{\pi}^3}})f({{m_q}\over {m_b}})
[1-{{2}\over {3}}{{{\alpha}_s}\over {\pi}}g({{m_q}\over {m_b}})],
\label{f1}
\end{equation}
where $m_q$ is the mass of the final $q$-quark decayed from $b$-quark.
As can be seen, the decay width of the spectator model depends on $m_b^5$,
therefore small difference of $m_b$ would change the decay width significantly.

Altarelli $et$ $al.$ \cite{alta} proposed for the inclusive $B$-meson
semileptonic decays their ACCMM model, which incorporates the bound state
effect by treating the $b$-quark as a virtual state particle, thus giving
momentum dependence to the $b$-quark mass.
The virtual state $b$-quark mass
$W$ is given by
\begin{equation}
W^2({\bf p})=m_B^2+m_{sp}^2-2m_B{\sqrt{{\bf p}^2+m_{sp}^2}}
\label{f2}
\end{equation}
in the $B$-meson rest frame, where $m_{sp}$ is the spectator quark mass,
$m_B$ is the $B$-meson mass, and {\bf p} is the momentum of
the $b$-quark inside $B$-meson.

For the momentum distribution of the virtual $b$-quark, Altarelli $et.$ $al.$
considered the Fermi motion inside the $B$-meson with the Gaussian momentum
distribution
\begin{equation}
\phi ({\bf p})={{4}\over {{\sqrt{\pi}}p_{_F}^3}}e^{-{\bf p}^2/p_{_F}^2},
\label{f3}
\end{equation}
where the Gaussian width, $p_{_F}$, is treated as a free parameter.
Then the lepton energy spectrum of the $B$-meson  decay is given by
\begin{equation}
{{d{\Gamma}_B}\over {dE_l}}(p_{_F}, m_{sp}, m_q, m_B)=
{\int}_0^{p_{max}}dp\ p^2\phi ({\bf p})\
{{d{\Gamma}_b}\over{dE_l}}(m_b=W, m_q),
\label{f4}
\end{equation}
where $p_{max}$ is the maximum kinematically allowed value of $p=|{\bf p}|$.
The ACCMM model, therefore, introduces a new parameter $p_{_F}$ for the
Gaussian momentum distribution of the $b$-quark
inside $B$-meson instead of the $b$-quark mass of the spectator model.
In this way the ACCMM model incorporates the bound state effects and reduces
the strong dependence on $b$-quark mass in the decay width of the
spectator model.

The Fermi momentum $p_{_F}$ is the most essential parameter of
the ACCMM model as we see in the above.
However, the experimental determination of its value
from the lepton energy spectrum has been very ambiguous, because
various parameters of ACCMM model, such as $p_{_F}$, $m_q$ and $m_{sp}$,
are fitted
all together from the limited region of end-point lepton energy spectrum,
and because the perturbative QCD corrections are very sensitive
in the end-point region of the spectrum. Recently, ARGUS \cite{argus2}
extracted the lepton energy spectrum of $B \rightarrow X_c l \nu$
for the whole region of electron energy, but
with much larger uncertainties. We argue that
the value $p_{_F} \sim 0.3$ GeV, which has been commonly
used in experimental analyses, has no theoretical or experimental clear
justification, even though there has been recently an assertion
that the prediction of heavy quark effective theory
approach~\cite{bigi}, far from the end-point region,
gives approximately equal shape to the ACCMM model with $p_{_F} \sim 0.3$ GeV.
Therefore, it is stongly recommended to determine the value of $p_{_F}$ more
reliably and independently,
when we think of the importance of its role in experimental analyses.
It is particularly important in the determination of the value of
$|V_{ub}/V_{cb}|$, as we explain in section 2.
A better
determination of $p_{_F}$ is also interesting theoretically since it has
its own physical correspondence related to the Fermi motion inside $B$-meson.
In this context we
calculate theoretically the value of $p_{_F}$ in the relativistic
quark model using quantum mechanical variational method in section 3.
And we obtain $p_{_F} = 0.54$ GeV which is much larger than 0.3 GeV.
Section 4 contains the conclusion.\\

\noindent
{\bf \large 2. Dependence of $|V_{ub}/V_{cb}|$
on the Fermi momentum parameter $p_{_F}$}\\

The ACCMM model provides an inclusive lepton energy spectrum of the $B$-meson
semileptonic decay to obtain the value of $|V_{ub}/V_{cb}|$. The
leptonic energy spectrum is useful in separating $b \rightarrow u$
transitions from $b \rightarrow c$,
since the end-point region of the spectrum is completely composed of
$b \rightarrow u$ decays. In applying this method one integrates
(\ref{f4}) in the range $2.3~{\rm GeV}<E_l<2.6~{\rm GeV}$ at
the $B$-meson rest frame, where only
$b \rightarrow u$ transitions exist \cite{cleo2}.
So we theoretically calculate
\begin{equation}
{\widetilde{\Gamma}}(p_{_F}) \equiv
\int_{2.3}^{2.6} dE_l
{{d{\Gamma}_B}\over {dE_l}}(p_{_F}, m_{sp}, m_q, m_B)~.
\label{z1}
\end{equation}
In (\ref{z1}) we specified only $p_{_F}$ dependence explicitly in the
left-hand side, and ${d{\Gamma}_B} / {dE_l}$ in the right-hand side is from
(\ref{f4}). Then one compares the theoretically calculated
${\widetilde{\Gamma}}(p_{_F})$ with the experimentally measured width
${\widetilde{\Gamma}}_{exp}$ in the region $2.3~{\rm GeV}<E_l<2.6~{\rm GeV}$,
to extract the value of $|V_{ub}|$ from the relation
\begin{equation}
{\widetilde{\Gamma}}_{exp}~=~|V_{ub}|^2~\times {\widetilde{\Gamma}}(p_{_F})~.
\label{z2}
\end{equation}
In the real experimental situation \cite{cleo,argus,argus2,cleo2},
the only measured quantity is the number of events in this region of
high $E_l$ compared to the total semileptonic events number,
{\it i.e.} the branching-fraction
${\widetilde{\Gamma}}_{exp} / {\widetilde{\Gamma}}^{total}_{s.l.}$.
Since the value ${\widetilde{\Gamma}}^{total}_{s.l.}$ is
propotional to $|V_{cb}|^2$, only the combination $|V_{ub} / V_{cb}|^2$ is
extracted.

We now consider the possible dependence of $|V_{ub} / V_{cb}|^2$
as a function of the
parameter $p_{_F}$ from the following relation
\begin{equation}
\frac{{\widetilde{\Gamma}}_{exp}} {{\widetilde{\Gamma}}^{total}_{s.l.}}~
\propto~
\left| \frac{V_{ub}}{V_{cb}} \right|^2_{p_{_F}=p_{_F}}
         \times {\widetilde{\Gamma}}(p_{_F})
=
\left| \frac{V_{ub}}{V_{cb}} \right|^2_{p_{_F}=0.3}
         \times {\widetilde{\Gamma}}(p_{_F}=0.3)~,
\label{z3}
\end{equation}
where
$|V_{ub}/V_{cb}|^2_{p_{_F}=p_{_F}}$ is
determined with an arbitrary value of the Fermi momentum parameter $p_{_F}$.
In the right-hand side we used $p_{_F}$=0.3 GeV because this value is
commonly used in the experimental determination of
$\left| {V_{ub}} / {V_{cb}} \right|$.
Then one can get a relation
\begin{equation}
\left| \frac{V_{ub}}{V_{cb}} \right|^2_{p_{_F}=p_{_F}}
=
\left| \frac{V_{ub}}{V_{cb}} \right|^2_{p_{_F}=0.3}
\times
\frac{\widetilde{\Gamma}(0.3)}{\widetilde{\Gamma}(p_{_F})}~.
\label{z4}
\end{equation}

In section 3, we obtain $p_{_F}=0.54$ GeV using the variational method in the
relativistic quark model.
If we use $p_{_F}=0.5$ GeV, instead of $p_{_F}=0.3$ GeV, in the experimental
analysis of the end-point region of lepton energy spectrum, the value of
$|V_{ub}/V_{cb}|$ becomes significantly changed.
We numerically calculated theoretical ratio
$\widetilde{\Gamma}(0.3)/\widetilde{\Gamma}(0.5)$
by using (\ref{f4}) and (\ref{z1}) with $m_{sp}=0.15$ GeV, $m_q=0.15$ GeV,
which are the values commonly used by experimentalists,
and $m_B=5.28$ GeV, to get its value as $1.81$, which finally gives
\begin{equation}
\left| \frac{V_{ub}}{V_{cb}} \right|^2_{p_{_F}=0.5}
=\left| \frac{V_{ub}}{V_{cb}} \right|^2_{p_{_F}=0.3} \times 1.81~.
\label{z5}
\end{equation}

Previously the CLEO \cite{cleo2} analyzed with
$p_{_F}=0.3$ GeV the end-point lepton energy spectrum to get
\begin{eqnarray}
10^2 \times |V_{ub}/V_{cb}|^2&=&0.57\pm 0.11 ~~({\rm ACCMM}~ \cite{alta})
\nonumber\\
&=&1.02\pm 0.20 ~~({\rm Isgur~et.al.}~ \cite{isgw})~.
\label{g3}
\end{eqnarray}
As can be seen, those values are in large disagreement.
However, if we use $p_{_F}=0.5$ GeV, the result of the ACCMM model
becomes $1.03$, and  these two models are in a good
agreement for the value of $|V_{ub}/V_{cb}|$.
Finally we show the values of $|V_{ub}(p_{_F})/V_{ub}(p_{_F}=0.3)|$ as
a function of $p_{_F}$ in Fig. 1.\\

\noindent
{\bf \large 3. Calculation of $p_{_F}$ in the relativistic quark model}\\

We consider the Gaussian probability distribution function $\phi ({\bf p})$
in (3) as
the absolute square of the momentum space wave function $\chi ({\bf p})$ of
the bound state $B$-meson, that is,
\begin{equation}
\phi ({\bf p}) = 4\pi \vert \chi ({\bf p}){\vert}^2,
\ \ \
\chi ({\bf p}) = {{1}\over {({\sqrt{\pi}}p_{_F})^{3/2}}}
e^{-{\bf p}^2/2p_{_F}^2}.
\label{f5}
\end{equation}
The Fourier transform of $\chi ({\bf p})$ gives the coordinate space wave
function $\psi ({\bf r})$, which is also Gaussian,
\begin{equation}
\psi ({\bf r})=({{p_{_F}}\over {\sqrt{\pi}}})^{3/2}e^{-r^2p_{_F}^2/2}.
\label{f7}
\end{equation}
Then we can approach the determination of $p_{_F}$ in the framework of
quantum mechanics.
For the $B$-meson system we treat the $b$-quark non-relativistically,
but the $u$- or $d$-quark relativistically with the Hamiltonian
\begin{equation}
H=M+{{{\bf p}^2}\over {2M}}+{\sqrt{{\bf p}^2+m^2}}+V(r),
\label{f8}
\end{equation}
where $M=m_b$ is the $b$-quark mass and $m=m_{sp}$ is the $u$- or $d$-quark
mass.
We apply the variational method to the Hamiltonian (\ref{f8})
with the trial wave function
\begin{equation}
\psi ({\bf r})=({{\mu}\over {\sqrt{\pi}}})^{3/2}e^{-{\mu}^2r^2/2},
\label{f9}
\end{equation}
where $\mu$ is the variational parameter.
The ground state is given by
minimizing the expectation value of $H$,
\begin{equation}
\langle H\rangle =\langle\psi\vert H\vert\psi\rangle =E(\mu ),
\ \ \
{{d}\over {d\mu }}E(\mu )=0\ \ {\rm{at}}\ \ \mu ={\bar{\mu}},
\label{f10}
\end{equation}
and then ${\bar{\mu}} = p_{_F}$ and $\bar E \equiv E({\bar{\mu}})$
approximates $m_B$.
The value of $\mu$ or $p_{_F}$ corresponds to the
measure of the radius of the two body bound state as can be seen from
$\langle r\rangle ={{2}\over{\sqrt{\pi}}}{{1}\over {\mu}}$
and
$\langle r^2{\rangle}^{{1}\over {2}} ={{3}\over {2}}{{1}\over {\mu}}$.

In (\ref{f8}) we take the Cornell potential
which is composed of the Coulomb and linear potentials,
\begin{equation}
V(r)=-{{{\alpha}_c}\over {r}}+Kr.
\label{f13}
\end{equation}
For the values of the parameters
${\alpha}_c\ (\equiv {{4}\over {3}}{\alpha}_s)$, $K$,
and the $b$-quark mass $m_b$, we use the values given
by Hagiwara $et$ $al.$~\cite{hagi},
\begin{equation}
{\alpha}_c=0.47\ ({\alpha}_s=0.35),\ \ K=0.19\ GeV^2,\ \ m_b=4.75\ GeV,
\label{f14}
\end{equation}
which have been determined by the best fit of all the known
$(c{\bar{c}})$ and $(b{\bar{b}})$ bound states.
For comparison we will also consider
${\alpha}_c=0.32\ ({\alpha}_s=0.24)$,
which corresponds to $\alpha_s(Q^2 = m_B^2)$.

Before applying our variational method with the Gaussian trial wave function
to the $B$-meson system, let us check the method by considering the
$\Upsilon(b{\bar{b}})$ system.
The Hamiltonian of the $\Upsilon (b{\bar{b}})$ system can be approximated
by the  non-relativistic Hamiltonian
\begin{equation}
H\simeq 2m_b+{{{\bf p}^2}\over {m_b}}+V(r).
\label{f16}
\end{equation}
With the parameters in (\ref{f14})
(or with ${\alpha}_c=0.32$),
our variational method with
the Gaussian trial wave function (\ref{f9})
gives $p_{_F}={\bar{\mu}}=1.1$ GeV and
${\bar{E}}=E({\bar{\mu}})=9.49$ GeV. Here $p_{_F}=1.1$ GeV corresponds to the
radius $R(\Upsilon )=0.2$ fm, and ${\bar{E}(\Upsilon)}=9.49$ GeV is within
$0.3\ \%$  error compared with the experimental value $E_{\rm{ exp}}=
m_{\Upsilon}=9.46$ GeV. Therefore, the
variational method with the non-relativistic Hamiltonian (\ref{f16})
gives fairly
accurate results for the $\Upsilon$ ground state.

However, since the $u$- or $d$-quark in the $B$-meson is very light, the
non-relativistic description can not be applied to the $B$-meson system.
For example, when we apply the variational method with the non-relativistic
Hamiltonian to the $B$-meson, we get the results
\begin{eqnarray}
p_{_F}=0.29\ GeV,\ \ {\bar{E}} =5.92\ GeV & & {\rm{for}}\ \
{\alpha}_s=0.35,
\label{f17}\\
p_{_F}=0.29\ GeV,\ \ {\bar{E}} =5.97\ GeV & & {\rm{for}}\ \
{\alpha}_s=0.24.
\label{f18}
\end{eqnarray}
The above masses $\bar E$ are much larger compared to the experimental value
$m_B=5.28$ GeV, and moreover the expectation values of the higher
terms in the non-relativistic perturbative expansion are bigger than those of
the lower terms. Therefore, we can not apply the variational method with the
non-relativistic Hamiltonian to the $B$-meson system.

Let us come back to our Hamiltonian (\ref{f8}) of the $B$-meson system.
In our variational method the trial wave
function is Gaussian both in the coordinate space and in the momentum space,
so the expectation value of $H$ can be calculated in either space from
$\langle H\rangle =\langle\psi({\bf r})\vert H\vert\psi({\bf r})\rangle
=\langle\chi({\bf p})\vert H\vert\chi({\bf p})\rangle$.
Also, the Gaussian function is a smooth function and its derivative of any
order is square integrable, thus any power of the Laplacian operator
${\nabla}^2$ is a hermitian operator at least under Gaussian functions.
Therefore, analyzing the Hamiltonian (\ref{f8}) with the variational method
can be considered as reasonable even though solving the eigenvalue equation
of the differential operator (\ref{f8}) may
be confronted with the mathematical difficulties because of the square root
operator in (\ref{f8}).

With the Gaussian trial wave function (\ref{f5}) or (\ref{f9}),
the expectation value of the Hamiltonian (\ref{f8}) can
be calculated easily besides the square root operator,
\begin{eqnarray}
\langle {\bf p}\,^2\rangle &=& \langle \psi ({\bf r}\,) |
{\bf p}\,^2| \psi ({\bf r}\,) \rangle =
\langle \chi ({\bf p}\,) | {\bf p}\,^2|
\chi ({\bf p}\,) \rangle = {3 \over 2} \mu^2,
\label{f21}\\
\langle V(r) \rangle &=& \langle \psi ({\bf r}) | -{\alpha_c \over r} + Kr \
|\psi ({\bf r}) \rangle = {2 \over \sqrt\pi} (-\alpha_c\mu + {K / \mu} ).
\label{f22}
\end{eqnarray}
Now let us consider the expectation value of the square root operator  in
the momentum space
\begin{eqnarray}
\langle \sqrt{{\bf p}\,^2+ m^2} \rangle &=& \langle \chi ({\bf p}\,)
| \sqrt{{\bf p}\,^2+ m^2} | \chi ({\bf p}\,) \rangle
= \Bigl({\mu \over \sqrt\pi}\Bigr)^3 \int_0^\infty
e^{-{p^2 / \mu^2}} \sqrt{{\bf p}\,^2+ m^2}\; d^3p
\nonumber\\
&=& {4\mu \over \sqrt\pi} \int_0^\infty e^{-x^2} \sqrt{x^2 + (m/\mu)^2} \;
x^2dx.
\label{f23}
\end{eqnarray}
The integral (\ref{f23}) can be given as a series expansion by the following
procedure. First, define
\begin{eqnarray}
I(s)\, &\equiv & \int_0^\infty \sqrt{x^2 + s} \; x^2 e^{-x^2} dx
= s^2 \int_0^\infty \sqrt{t^2 + 1} \; t^2 e^{-st^2} dt,
\label{f24}\\
I_0(s) &\equiv & \int_0^\infty \sqrt{x^2 + s} \; e^{-x^2} dx
= s \int_0^\infty \sqrt{t^2 + 1} \; e^{-st^2} dt.
\label{f25}
\end{eqnarray}
Next,  from (\ref{f24}) and (\ref{f25}),
we find the following differential relations
\begin{equation}
{d \over ds} \Bigl({I_0 \over s}\Bigr) = - {1 \over s^2} I ,
\ \ \
{d I \over d s} = - {1 \over 2} I_0 + I .
\label{f26}
\end{equation}
Combining two equations in (\ref{f26}),
we get a second order differential equation for $I(s)$,
\begin{equation}
s I''(s) - (1+s) I'(s) + {1 \over 2} I(s) = 0.
\label{f28}
\end{equation}
The series solution to (\ref{f28}) is given as
\begin{eqnarray}
I(s) &=& c_1 I_1 (s) + c_2 I_2 (s),
\nonumber\\
I_1 (s) &=& s^2 F(s; {3 \over 2}, 3) = s^2 \Bigl\{ 1 + {1 \over 2} s +
{5 \over 32} s^2 + {7 \over 192} s^3 + {7 \over 1024} s^4 + \cdots\Bigr\},
\label{f29}\\
I_2 (s) &=& I_1 (s) \int {s e^s \over [I_1 (s)]^2} ds
= - {1 \over 16} s^2 \ln s \Bigl( 1 + {1 \over 2} s + {5 \over 32} s^2 +
\cdots \Bigr)
\nonumber\\
&-&{1 \over 2} \Bigl( 1 + {1 \over 2} s + {5 \over 32} s^2 +
{7 \over 192} s^3 + {7 \over 1536} s^4 + \cdots \Bigr),
\nonumber
\end{eqnarray}
where $F(s; {3\over 2}, 3)$ is the confluent hypergeometric function which
is convergent for any finite $s$, and the integral constants
$c_1\simeq-0.095$,  $c_2=-1$.
See  Appendix for the derivation of these numerical values for $c_i$.

Finally, collecting (\ref{f21}), (\ref{f22}) and (\ref{f23}),
the expectation value of $H$ is written as
\begin{eqnarray}
\langle H \rangle &=& M + {1\over 2M} \Bigl({3\over 2} \mu^2
\Bigr) + {2 \over \sqrt\pi} ( -\alpha_c \mu + K/\mu )
\nonumber\\
& & + {2\mu \over \sqrt\pi} \biggl[ 1 + {1\over 2} (m/\mu)^2 +
\Bigl({5\over 32} - 2c_1 \Bigr) (m/\mu)^4 + {1\over 4} (m/\mu)^4 \ln(m/\mu)
\biggr] ,
\label{f30}
\end{eqnarray}
up to $(m/\mu)^4$.

With  the input value of $m =m_{sp} = 0.15$ GeV, we minimize
$\langle H \rangle$ of (\ref{f30}), and then we obtain
\begin{eqnarray}
p_{_F}=\bar \mu &=& 0.54 \ GeV, \qquad m_B=\bar E = 5.54 \ GeV
\qquad {\rm for}\;
\alpha_s=0.35,
\label{f31}\\
\bar \mu &=& 0.49 \ GeV, \ \ \qquad\qquad\bar E = 5.63 \ GeV \ \qquad
{\rm for}\; \alpha_s=0.24.
\nonumber
\end{eqnarray}
Here let us check how much sensitive our calculation of $p_{_F}$ is
by considering the case where $m=m_{sp}=0$ for comparison.
For $m_{sp}=0$ the integral in (\ref{f23}) is done easily and we obtain the
following values of $\bar \mu =p_{_F}$ by the above variational method.
\begin{eqnarray}
\bar \mu &=& 0.53 \ GeV, \qquad\bar E = 5.52 \ GeV \qquad {\rm for}\;
\alpha_s=0.35,
\label{f32}\\
\bar \mu &=& 0.48 \ GeV, \qquad\bar E = 5.60 \ GeV \qquad {\rm for}\;
\alpha_s=0.24.
\nonumber
\end{eqnarray}
As we see in (\ref{f32}), the results are similar to those in (\ref{f31})
where
$m_{sp}=0.15$ GeV.
We could expect this insensitivity of the value of $p_{_F}$ to that of
$m_{sp}$ because the value of $m_{sp}$, which should be small
in any case, can not affect the integral in (23) significantly.

The calculated values of the $B$-meson mass, $\bar E$,  are much larger than
the measured  value of 5.28 GeV.
The large values for the mass are   originated  partly because the Hamiltonian
(\ref{f30})  does not take care of the correct spin dependences
for $B$ and $B^*$.
The difference between the pseudoscalar meson and the vector meson is  given
arise to by the
chromomagnetic hyperfine splitting, which is given by 
\begin{equation}
V_s = {2 \over 3Mm} \; \vec s_1 \cdot \vec s_2 \,\nabla^2
(- {\alpha_c \over r}).
\label{f33}
\end{equation}
Then the expectation values of $V_s$ are given by
\begin{equation}
\langle V_s \rangle = - {2 \over \sqrt\pi} \,
{\alpha_c \mu^3 \over Mm} \ \ {\rm for} \ \ B,
\ \ \
\langle V_s \rangle
= {2 \over 3\sqrt\pi} \, {\alpha_c \mu^3 \over Mm} \ \ {\rm for}
\ \ B^*,
\label{f34}
\end{equation}
and we treat $\langle V_s \rangle$ only as a perturbation.
Then with  the input value of $m =m_{sp} = 0.15$ GeV, we get for $B$ meson
\begin{eqnarray}
p_{_F} &=& 0.54 \ GeV, \qquad \bar E_B = 5.42 \ GeV \qquad
{\rm for}\; \alpha_s=0.35,
\label{f35}\\
p_{_F} &=& 0.49 \ GeV, \qquad \bar E_B = 5.56 \ GeV\qquad
{\rm for}\; \alpha_s=0.24 ,
\nonumber
\end{eqnarray}
and for $B^*$
\begin{eqnarray}
p_{_F} &=& 0.54 \ GeV, \qquad \bar E_{B^*} = 5.58 \ GeV \qquad
{\rm for}\; \alpha_s=0.35,
\label{f36}\\
p_{_F} &=& 0.49 \ GeV, \qquad \bar E_{B^*} = 5.65 \ GeV\qquad
{\rm for}\; \alpha_s=0.24.
\nonumber
\end{eqnarray}

The calculated values of the $B$-meson mass, 5.42 GeV ($\alpha_s = 0.35$) and
5.56 GeV ($\alpha_s = 0.24$) are in reasonable agreement compared to the
experimental value of $m_B=5.28$ GeV; the relative errors are 2.7\% and 5.3\%,
respectively.
However, for the Fermi momentum $p_{_F}$, the calculated values, 0.54 GeV
($\alpha_s = 0.35$) and  0.49 GeV
($\alpha_s = 0.24$), are larger than the value 0.3 GeV,
which has been commonly used
in the experimental
analyses of energy spectrum of semileptonic $B$-meson decay.
The value $p_{_F} = 0.3$ GeV corresponds to the $B$-meson radius
$R_B \sim 0.66$ fm,  which
seems too large.
On the other hand, the value $p_{_F} = 0.5$ GeV corresponds to
$R_B \sim 0.39$ fm, which looks  in reasonable range.\\

\noindent
{\bf \large 4. Conclusion}\\

The Gaussian width of Fermi motion, $p_{_F}$, is the most
important parameter of the
ACCMM model, and the value $p_{_F} \sim 0.3$ GeV has been commonly used in
experimental  analyses without clear theoretical or experimental evidence.
Therefore, it is recommended to determine the value of $p_{_F}$ more
reliably, when we think of its importance in experimental analyses.
We calculated the
value for $p_{_F}$ in the relativistic quark model using the
variational method.
We obtained $p_{_F} = 0.54$ GeV, which is much
larger than 0.3 GeV. We also derived the ground state eigenvalue of
$E_B \simeq 5.5$ GeV, which is in reasonable agreement with
the experimental value $m_B=5.28$ GeV.

We studied the dependence of $\left| {V_{ub}} / {V_{cb}} \right|$
on the Fermi momentum parameter $p_{_F}$ in the ACCMM model, and extracted
$\left| {V_{ub}} / {V_{cb}} \right|$ as a function of $p_{_F}$.
It shows that $\left| {V_{ub}} / {V_{cb}} \right|$ is very much dependent on
the value of $p_{_F}$.
When we use $p_{_F} = 0.5$ GeV instead of 0.3 GeV,
$|V_{ub}/V_{cb}|$ is increased by a factor 1.81.
Then the previous discrepancy between the ACCMM model and the Isgur $et$ $al.$
model for the value of $\left| {V_{ub}} / {V_{cb}} \right|^2$ turns into a
good agreement.\\

\noindent
{\em Acknowledgements} \\
\indent
The work  was supported
in part by the Korean Science and Engineering  Foundation,
Project No. 931-0200-021-2, 951-0207-008-2,
in part by Non-Directed-Research-Fund, Korea Research Foundation 1993,
in part by the Center for Theoretical Physics, Seoul National University,
in part by Yonsei University Faculty Research Grant,
in part by Dae Yang Academic Foundation,  and
in part by the Basic Science Research Institute Program,
Ministry of Education 1994,  Project No. BSRI-94-2425.

\vspace{1.0cm}

After we submitted this paper, we have been informed an interesting
work by C. Greub {\it et. al.} \cite{greub},
which reports a similar conclusion
for the value of $p_{_F}$, that is $p_{_F}=566$ MeV, 
even though they used totally different approach.

\pagebreak

\centerline{\bf Appendix}

\medskip

The integration constants $c_1$ and $c_2$ in (\ref{f29}) are given
by the following
relations,
\begin{eqnarray}
I(0) &=& -{1\over 2}c_2 = \int_0^\infty x^3e^{-x^2}dx = {1\over 2},
\label{a1}\\
I''(s\approx 0) &=& 2c_1 + c_2 (-{1\over 8}\ln s -{11\over 32})
\nonumber\\
&=& -{1\over 4} \int_0^\infty x^2(x^2+s)^{-3/2}e^{-x^2}dx \quad {\rm at}
\quad s\approx 0 .
\label{a2}
\end{eqnarray}
Then, from (\ref{a1}), we get
\begin{equation}
c_2\ \ =\ \ -1.
\label{a3}
\end{equation}
The integral in (\ref{a2}) can be expanded as
\begin{eqnarray}
J(s=a^2) &=&\int_0^\infty x^2(x^2+a^2)^{-3/2}e^{-x^2}dx
\nonumber\\
&=& \int_0^\infty x^2\big[(x+a)^2-2ax\big]^{-3/2}e^{-x^2}dx
\nonumber\\
&=& \int_0^\infty x^2(x+a)^{-3}\left[ 1-{2ax\over (x+a)^2} \right]^{-3/2}
e^{-x^2}dx
\nonumber\\
&=& \sum_{n=0}^\infty {(2n+1)!a^n \over 2^n(n!)^2}\int_0^\infty {x^{n+2}\over
(x+a)^{2n+3}}e^{-x^2}dx.
\label{a4}
\end{eqnarray}
Next the integral in (\ref{a4}) is obtained by
\begin{equation}
\int_0^\infty {x^{n+2}\over (x+a)^{2n+3}}e^{-x^2}dx = {1\over (2n+2)!}
\left({\partial\over\partial a}\right)^{2n+2}
\int_0^\infty {x^{n+2}\over x+a}e^{-x^2}dx.
\label{a5}
\end{equation}
Again the integral in (\ref{a5}) is related to another integral
\begin{equation}
\int_0^\infty{x^{n+2}\over x+a}e^{-x^2}dx = \sum_{k=0}^{n+1}{(-a)^k \over
2} \left({n-k\over 2}\right)!+(-a)^{n+2}\int_0^\infty{e^{-x^2}\over x+a}dx.
\label{a6}
\end{equation}
The integral in (\ref{a6}) is given, for an infinitesimal value of $a$, by
integration by parts,
\begin{equation}
\int_0^\infty {e^{-x^2}\over x+a} dx =
e^{-x^2}~ln(x+a){\vert^{\infty}_0} -
\int_0^\infty e^{-x^2} (-2x) ln(x+a) dx =
-ln~a - {\frac{\gamma}{2}} + O(a),
\label{y1}
\end{equation}
where $\gamma \sim 0.5772$ is the Euler's constant.
Collecting (\ref{a5}), (\ref{a6}), and (\ref{y1}),
\begin{equation}
J(a\approx 0) =
\sum_{n=0}^\infty {1 \over 2^n(n!)^2 (2n+2)} a^n
\left({\partial\over\partial a}\right)^{2n+2}
(-a)^{n+2}
\{ -ln~a - {\frac{\gamma}{2}} + O(a) \}.
\label{y2}
\end{equation}
To get the constant $c_1$, we should extract a logarithmic term and constants
from (\ref{y2}),
\begin{equation}
J(a\approx 0) =
( -ln~a - {\frac{\gamma}{2}} - {3\over 2} ) +
\sum_{n=1}^\infty {(-1)^{n+1} \over 2^n(n!)^2 (2n+2)} a^n
\left({\partial\over\partial a}\right)^{2n+2}
( a^{n+2} ln~a ),
\label{y3}
\end{equation}
\begin{eqnarray}
a^n \left({\partial\over\partial a}\right)^{2n+2} ( a^{n+2} ln~a )
\vert_{a=0}
&=& \sum_{k=0}^{n+2} {(2n+2)! \over k! (2n+2-k)!}
{(n+2)! (2n+1-k)! (-1)^{k+1} \over (n+2-k)!}
\nonumber\\
&=& (-1)^{n+1} (n+2)! (n-1)! .
\label{y4}
\end{eqnarray}
Inserting (\ref{y4}) into (\ref{y3}), we get
\begin{equation}
J(a\approx 0) =
 -ln~a - {\frac{\gamma}{2}} - 1 + \beta ,
\label{y5}
\end{equation}
where
$\beta = \sum_{n=1}^{\infty} 1 / (n 2^n) \approx 0.6932$.
Then, from (\ref{a2}), we get
\begin{equation}
c_1 = -{3\over 64} + {\gamma \over 16} - {1\over 8} \beta ~
\approx -0.0975 .
\label{a8}
\end{equation}

\pagebreak

\pagebreak

\vspace*{3.5cm}
\vspace*{14.5cm}
\noindent
Fig.1 The ratio  $|V_{ub}(p_{_F})/V_{ub}(p_{_F}=0.3)|$ as
a function of $p_{_F}$.

\end{document}